\documentclass[12pt]{article}\usepackage[hyperfootnotes=false]{hyperref}
\usepackage{epsfig}
\usepackage{float}
\usepackage{dsfont}
\usepackage{comment}
\usepackage{bbold}
\usepackage{color}
\usepackage{caption}
\usepackage{subcaption}
\usepackage{mciteplus} 
\usepackage{amsmath}
\usepackage{amssymb}
\usepackage{graphicx}
\newlength{\abstractwidth}
\renewcommand{\thefootnote}{\fnsymbol{footnote}}
\renewcommand{\thanks}[1]{\footnote{#1}} 
\newcommand{\starttext}{
\setcounter{footnote}{0}
\renewcommand{\thefootnote}{\arabic{footnote}}}

\newcommand{\be}{\begin{equation}}
\newcommand{\bea}{\begin{eqnarray}}
\newcommand{\eea}{\end{eqnarray}}
\newcommand{\beq}{\begin{equation}}
\newcommand{\ee}{\end{equation}}
\newcommand{\C}{\mathbb{C}}

\DeclareMathOperator{\Tr}{Tr}

\def\lb{\label}
\def\simleq{\; \raise0.3ex\hbox{$<$\kern-0.75em
\raise-1.1ex\hbox{$\sim$}}\; }
\def\simgeq{\; \raise0.3ex\hbox{$>$\kern-0.75em
\raise-1.1ex\hbox{$\sim$}}\; }
\def\s{$\sigma$}

\makeatletter
\g@addto@macro\normalsize{%
  \setlength\abovedisplayskip{10pt}
  \setlength\belowdisplayskip{20pt}
  \setlength\abovedisplayshortskip{10pt}
  \setlength\belowdisplayshortskip{20pt}
}
\makeatother

\def\bn{\bigskip \noindent}

\renewcommand{\title}[1]{\vbox{\center\LARGE{#1}}\vspace{5mm}}
\renewcommand{\author}[1]{\vbox{\center#1}\vspace{5mm}}
\newcommand{\address}[1]{\vbox{\center\em#1}}

\begin{document}
  
\begin{titlepage}

\rightline{}
\bigskip
\bigskip\bigskip\bigskip\bigskip
\bigskip

\centerline{\Large \bf {Cayley graphs and complexity geometry}}

\bn

\bigskip

\bigskip
\begin{center}

\author{Henry W. Lin$^{a,b}$}

\address{

$^{a}$ Jadwin Hall, Princeton University,\\
 Princeton, NJ 08540, USA\\
 \vspace{10pt}

$^{b}$ Facebook AI Research, Facebook,\\ New York, NY 10003, USA
}
\end{center}

\begin{center}
\bf     \rm

\bigskip

\end{center}

\begin{abstract}

The basic idea of quantum complexity geometry is to endow the space of unitary matrices with a metric, engineered to make complex operators far from the origin, and simple operators near. By restricting our attention to a finite subgroup of the unitary group, we observe that this idea can be made rigorous: the complexity geometry becomes what is known as a Cayley graph. 
This connection allows us to translate results from the geometrical group theory literature into statements about complexity. For example, the notion of $\delta$-hyperbolicity makes precise the idea that complexity geometry is negatively curved. We report an exact (in the large $N$ limit) computation of the average complexity as a function of time in a random circuit model.

\medskip
\noindent
\end{abstract}

\end{titlepage}

\starttext \baselineskip=17.63pt \setcounter{footnote}{0}


\tableofcontents

\newcommand{\E}[1]{\times 10^{#1}}
\newcommand{\lp}{\left (}
\newcommand{\rp}{\right )}
\newcommand{\rb}{\right ]}
\newcommand{\RA}{\Rightarrow}
\newcommand{\R}{\mathbb{R}}
\newcommand{\h}{\hbar}
\newcommand{\pa}[2]{\frac{\partial #1}{\partial #2}}
\newcommand{\pd}{\partial}
\newcommand{\eqn}[1]{\begin{equation}\begin{split} #1 \end{split}\end{equation}}
\newcommand{\bra}[1]{\left \langle #1\right|}
\newcommand{\ket}[1]{\left | #1 \right \rangle}
\newcommand{\cre}{a^\dagger}
\newcommand{\ts}{\otimes}
\newcommand{\ev}[1]{\left \langle #1 \right \rangle}

\newcommand{\grad}{\vec{\nabla}}
\newcommand{\rt}{\sqrt{2}}
\newcommand{\intl}[3]{\int_{#1}^{#2} \text{d} #3 \, }
\newcommand{\intn}[1]{\int \text{d} #1 \, }
\newcommand{\dd}[1]{\delta\lp #1 \rp }
\newcommand{\rd}{\text{d}}
\newcommand{\mpl}{M_\text{Pl}}
\newcommand{\M}{\mathcal{M}}
\newcommand{\dk}[1]{\frac{\text{d}^4 k_{#1}}{(2\pi)^4}}
\newcommand{\prop}[1]{\lp #1\rp^2 - m^2 + i \epsilon}
\newcommand{\propn}[1]{\lp #1\rp^2 + i \epsilon}
\newcommand{\dcross}{\frac{d\sigma}{d\Omega}}
\newcommand{\p}{\s{p}}
\newcommand{\sq}{\s{q}}
\newcommand{\ac}[2]{\left\{#1,#2\right\}}
\def\beqa#1{\begin{eqnarray}\label{#1}}
\def\eeqa{\end{eqnarray}}
\newcommand{\inv}{^{-1}}
\newcommand{\tr}[1]{\text{tr} \lb #1 \rb}

\section{Introduction}
One challenge of any holographic proposal relating the complexity of a boundary system with a bulk geometric quantity \cite{stanford2014complexity, brown2016holographic, brown2016complexity, susskind-firewall} is the severe difficulty of independently computing the complexity of any boundary system --- even one with a finite dimensional Hilbert space such as SYK \cite{syk}. Faced with such a challenge, it is worth asking if any toy model exists where complexity can be better understood. In this paper, we study a toy model where one can actually compute the complexity and see that it grows linearly for an exponentially long amount of time; this matches the linear growth of a wormhole.

Of course, a better understanding of quantum complexity is interesting in its own right. As has been emphasized recently \cite{nielsen2006quantum, brown17, susskind-lectures}, quantum complexity has a geometric character; if nothing else, this paper expounds on this and shows how mathematical tools from geometric group theory and related subjects can be fruitfully applied to analyze complexity. It seems possible that such tools could also be used to shed light on more traditional problems in complexity theory.

We begin by briefly reviewing one of the garden variety definitions of complexity usually called circuit complexity. It is defined as follows \cite{preskill1998lecture, susskind-lectures}.

Given a Hilbert space $\mathcal{H}$ corresponding to a physical system of interest, denote by $S$ a subset of the unitary operators acting on $\mathcal{H}$ that are ``simple'' operators or ``gates''. We imagine that these are the unitaries that are easy for a quantum mechanic to implement. Our main assumption about the subset $S$ is that it is sufficiently rich to ensure that any unitary can be decomposed into a product of simple operators to arbitrary accuracy. In addition, it will be convenient to assume that $S$ is closed under inverses $S = S\inv$. For example, if $\mathcal{H}$ has a tensor product factorization into qubits, $S$ could be the set of all operators which act only on $k$-qubits. 
Then given any unitary operator on $\mathcal{H}$, decompose it as a product of gates $U = s_1 \cdots s_
\ell$, where $s_i\in S$. The complexity of $U$ is the minimum such $\ell$. If $S$ is a finite set, then it will almost always be impossible to decompose $U$ into a finite product of $s$. So for finite $S$, we must loosen the criteria slightly by demanding an approximate decomposition of $U$. The tolerance $\epsilon$ (as measured by the inner product) may be considered an additional parameter of the circuit complexity $C = C_\epsilon$.

When encountering a new mathematical object, it is often useful to list its important properties, and then reverse the logic by considering these abstract properties as defining the object more generally. Let us therefore list some properties of $C$. If we define the relative complexity between two operators $d(U_1,U_2) = C(U_1 U_2\inv)$, then $d$ satisfies the properties of a metric: symmetry follows from $S= S\inv$, and the triangle inequality follows from composing the two circuits\footnote{Up to a subtlety involving the tolerance $\epsilon$.}.

Another important property is that for many choices of $S$, the maximum value of the complexity is exponentially large in $K$. On the other hand, the number of unitaries which differ from each other by more than $\epsilon$ as measured by the inner product is doubly exponential in $K$. In a geometric language, the volume of the space is exponential in the diameter. This is a strong hint that the complexity geometry should be negatively curved; in flat space, volume grows polynomially with the diameter; it grows even slower in spaces with positive curvature. Additional evidence that the geometry should be negatively curved can be found in \cite{brown16}.

The idea of complexity geometry \cite{nielsen2006quantum, brown17}, then, is to consider these properties as the defining ones for complexity. One searches for a smooth metric on the unitary group such that has the above properties. Of course, we do not expect that these properties uniquely specify a metric. Roughly speaking, for different choices of a simple set $S$, we will get different metrics. There are also somewhat arbitrary choices like requiring diagonal elements of the metric to vanish and fixing the functional form of the diagonal elements \cite{brown17}. In general, these choices yield metrics that are not related by diffeomorphisms; they represent genuinely different geometries.

The purpose of this paper is to put these ideas on a somewhat firmer footing, by considering the simplified setting of a finite (or at least discrete) subgroup of the unitary group. In $\S 2$, we discuss this setting and its connection to discrete mathematics. By making use of facts about the permutation group, we report an exact (in the large-$K$ limit) expression for the growth of complexity as a function of time in a toy model. In $\S 3$, we give some implications for complexity geometry in general and speculate about implications for holography. We have tried to write this paper for a diverse audience, so we assume minimal familiarity or interest in holography except in \S 3.

\section{Complexity geometry of a discrete subgroup}
Many of the above comments can be made rigorous in a simplified setting. The setting is to consider a finite but large subgroup of the unitary group.
By large, we mean that if we specialize to a system of Hilbert space dimension $D$, the order of the subgroup is still exponentially large in $D$. In the above definition of circuit complexity, we already implicitly considered a large but finite subset of the unitary group, defined so that there is one group element per $\epsilon$ ball. The simplification here comes from the assumption that this subset still has a group structure.

An interesting example of such a subgroup is the set of permutations on $2^K$ elements. These are exactly the reversible classical operations that can be performed on {\it bit strings} of length $K$. The order of this group is $(2^K)!$, which is a doubly exponentially large in $K$. Let us pause to emphasize that this is {\it not} the permutation group acting on individual bits, which is a much smaller group of order $K!$. For example, the CNOT gate is contained in the permutation group we are considering, but is not contained in the permutation group acting on bits.

 This group is interesting in its own right, independent from the above motivations. The connection to classical computation is discussed in the appendices. If we allow the elements of the permutation matrices to include $e^{2\pi i/p}$ for some integer $p$ instead of just 0 and 1, we get another example of a large finite subgroup which contains more than just classical operations. In fact, the qualification that a finite group be a subgroup of the unitary group is without content since any finite group has a faithful unitary representation. 

Finally, we will mention in passing that one could also consider a different toy model where one considers the unitary matrices over not the complex numbers but over a finite field. This is a sort of finite deformation of quantum mechanics, which might be interesting to study further.

Now we come to a main point. A group $G$ equipped with a finite generating set $S = S\inv$ has a natural graph structure known as a {\it Cayley graph}: the vertices of the graph are elements of $G$ and there is an edge between two vertices $g,h$ iff $gh\inv \in \mathcal{G}$. The Cayley graph induces a natural metric on $G$. The distance between two points in the graph is the minimal number of edges needed to connect the points. Since a geodesic connecting the identity and an element $g$ corresponds to a minimal decomposition $g = s_1 \cdots s_n$, with $s \in S$, this is a natural and precise definition of complexity geometry in a finite setting.\footnote{This point of view could be summarized with the slogan $CG = CG$, or Cayley graphs = complexity geometry.} The fact that there could be many generating sets $S$ corresponds to many different choices of simple gates.

Here is a simple example of a Cayley graph. If our Hilbert space is a single qubit, we can write any unitary as $U = e^{-i \vec{v} \cdot \vec{\sigma}}$ up to an overall phase. Let us suppose that our simple operations are $R_x(2\pi/3)$ and $R_y(2\pi/3)$, corresponding to $\vec{v} = \frac{2 \pi }{3} e_x$ and $\vec{v} = \frac{2 \pi }{3} e_y$. If we only consider either of these two generators, the generated subgroup of $SU(2)$ would just be $\mathbb{Z}_3$. One might think that with both generators, the resulting subgroup is $\mathbb{Z}_3 \times \mathbb{Z}_3$. This is incorrect since $R_x R_y \ne R_y R_x$. Instead, the generated subgroup is $G \cong \mathbb{Z}_3 \star \mathbb{Z}_3$, where $\star$ is the free product of two groups. The resulting Cayley graph looks like this\footnote{Figure from \url{https://commons.wikimedia.org/wiki/File:H2_tiling_33i-3.png}}:

\begin{figure}[H]
	\centering
	\includegraphics[scale=0.25, trim = 0cm 0cm 0cm 0cm, clip=true]{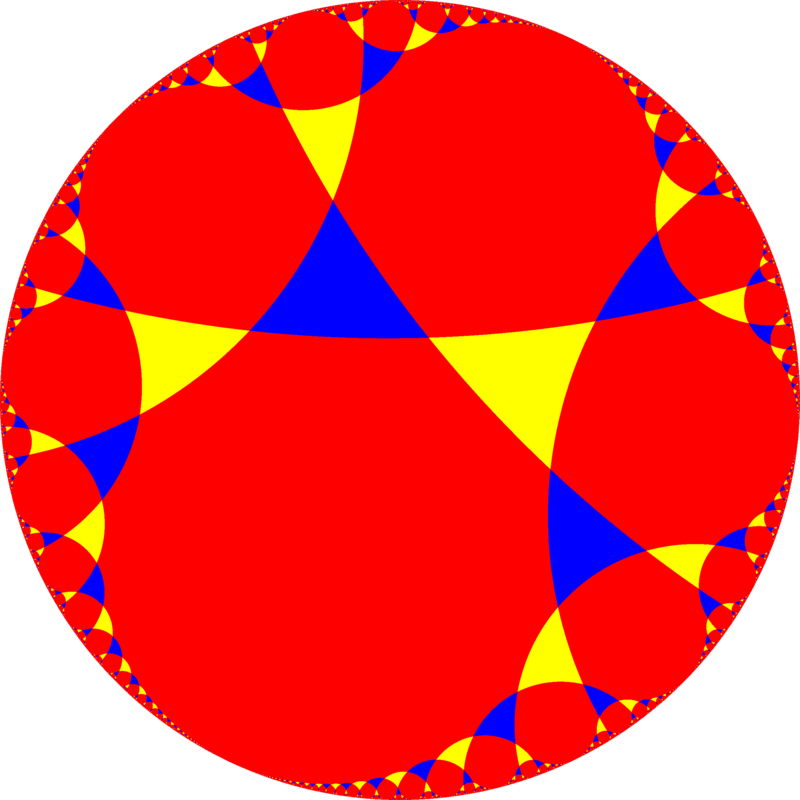}
	\caption[]{Cayley graph of $\mathbb{Z}_3 \star \mathbb{Z}_3$. The elements of the group live on the vertices of the triangles; the edges of the Cayley graph live on the edges of the triangles. Going around a blue (yellow) triangle clockwise corresponds to multiplication by $R_x$ ($R_y$). Going around a blue (yellow) triangle counterclockwise corresponds to multiplication by $R_x\inv$ ($R_y\inv$). The triangles have three sides since $R_x^3 = R_y^3 = 1$. This Cayley graph happens to give a uniform tiling of the hyperbolic plane, which shows that $\mathbb{Z}_3 \star \mathbb{Z}_3$ is $\delta$-hyperbolic.} 
	\label{fz3}
\end{figure}

Given a metric on some space, it is natural to wonder whether the space is curved or flat in some sense. Our first instinct might be to try to embed the graph in a smooth space where Riemannian notions of curvature apply. Indeed, in the above figure we have isometrically embedded the Cayley graph of $\mathbb{Z}_3 \star \mathbb{Z}_3$ in the Poincare plane, which suggests that there is some sense in which the Cayley graph is negatively curved. However, embedding a graph into a smooth geometry is typically a complicated thing to do. Fortunately, there is a simple notion of negative curvature due to Gromov that only relies on the intrinsic geometry.
 
In any metric space, one can define a triangle as three points joined by geodesics. A triangle is $\delta$-thin if for any point on the triangle, we can find a point on one of the other two sides of the triangle at a distance smaller than $\delta$. Intuitively, this says that the center of the triangle is not far from the edges. We say that a metric space is $\delta$-hyperbolic if all triangles\footnote{If we want a local notion of curvature, we just need to demand that triangles that are small compared to the diameter of the space are $\delta$-thin.} are $\delta$-thin. In a smooth hyperbolic space like the Poincare disk, $\delta$ is nothing but the curvature scale.

In a finite metric space, one can always trivially bound $\delta$ by the diameter of the space, which is the maximum distance between two points. In a space where the diameter is infinite, the condition of $\delta$-hyperbolicity is non-trivial. A finitely generated infinite group $G$ whose Cayley graph is $\delta$-hyperbolic is known as a {\it hyperbolic group}. The condition that $G$ be finitely generated just means that $S$ is finite. Of course, there could be many different $S$ which generate $G$. One might think that whether or not the Cayley graph is $\delta$-thin depends on the choice of generating set. A basic fact about hyperbolic groups is that while the value of $\delta$ may depend on $S$, the existence of a finite $\delta$ does not. 

Returning to the case of a finite group, if one considers a sequence of finite spaces $M_n$, one can ask what the minimum $\delta_n$ is for each of the spaces. If the diameter $D_n$ of $M_n$ grows without bound, the condition that $\delta_n < \delta_{\infty}$ is a non-trivial condition. In general, as long as $\delta_n/D_n \to 0$ as $n\to \infty$, there is a meaningful sense in which the sequence of finite groups is hyperbolic\footnote{For $K$ qubits, we need a curvature scale $\sim K$ to reproduce fast scrambling whereas the diameter is exponential in $K$. So we expect that $\delta$ can grow without bound, but still be a very small fraction of the diameter. }.

\begin{figure}[H]
	\centering
	\includegraphics[scale=0.33, trim = 0cm 0cm 0cm 0cm, clip=true]{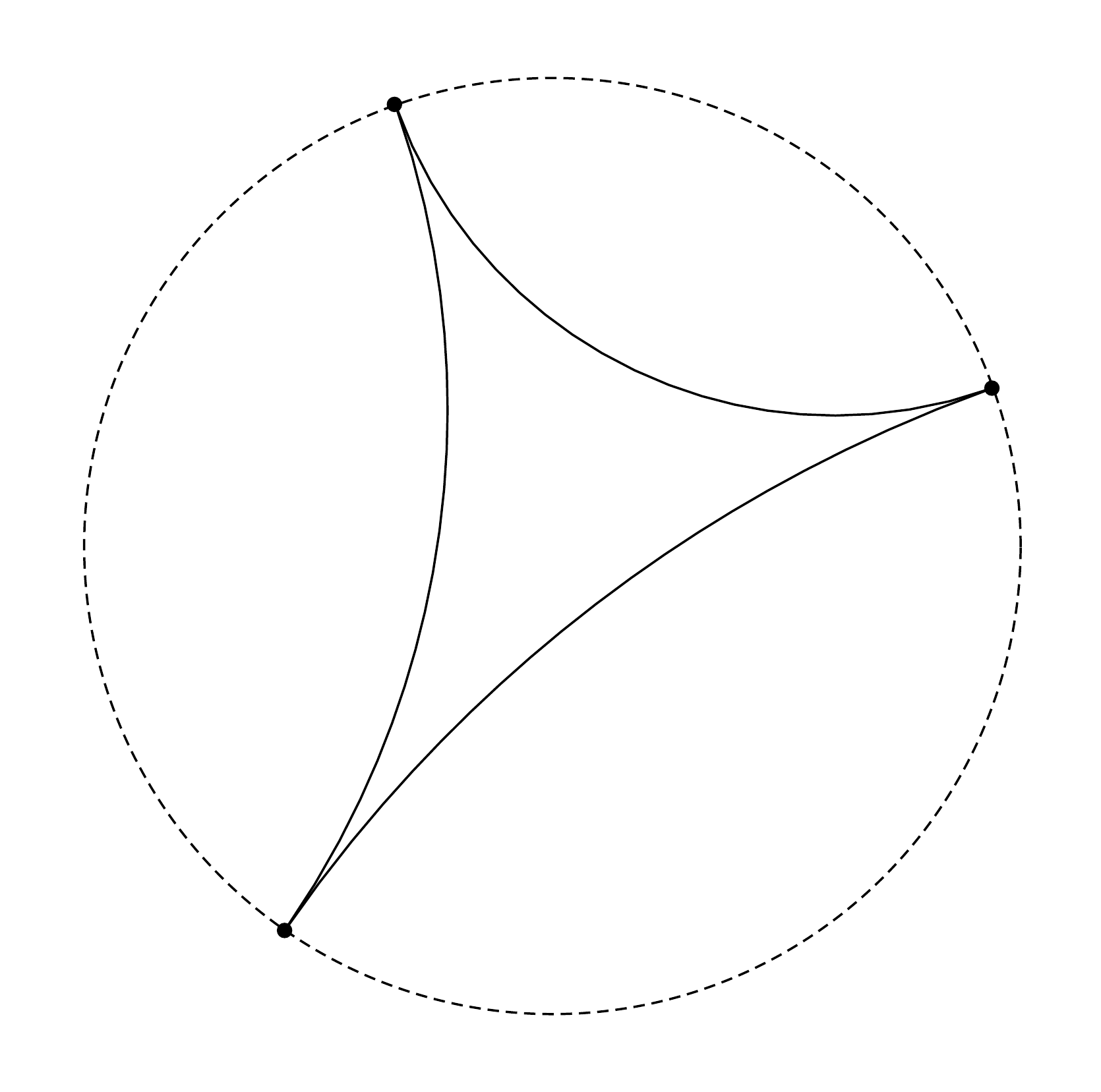}
	\caption{A large triangle on the Poincare disk. Even though the distance between the three points grows without bound, any point on one side of the triangle is close to some other point on another side of the triangle. The triangle is $\delta$-thin, where $\delta$ is of order the curvature scale.}
\end{figure}

\subsection{Permutation group}
We will now focus on the permutation subgroup $S_{n}$ with $n=2^{K-1}$. These are the set of all classical reversible computations that fix the last bit.
	A more detailed understanding of the complexity geometry is possible thanks to results in the math literature. The reason for choosing $n=2^{K-1}$ instead of $n=2^K$ is rather technical; we relegate its explanation to Appendix A. 
For $S$, we will choose the set of all transpositions. We emphasize again that the permutation group we are interested in does {\it not} act on $K-1$ bits but on the $2^{K-1}$ bit strings; these simple operations fix all but two bit strings. From the usual circuit point of view, this is a somewhat unusual generating set since transpositions are not $k$-local. Nevertheless, any transposition can be constructed from $k$-local gates with circuit complexity $\sim K$.
 So even if we take as simple the $k$-local gates, the transpositions are still relatively simple. Apart from its interest as a model of reversible classical computation, we have chosen this subgroup and this generating set so as to maximally connect with the math literature. No doubt many other subgroups could be studied in future work.

We will now list some facts about this geometry. The first one is elementary: the diameter of the Cayley graph is exactly $D_n = n-1$. To see this, note that any permutation can be decomposed uniquely into disjoint cycles. Now any cycle of length $k$ can be written as a product of $k-1$ transpositions. So $D_n \le n-1$. To show that $D_n \ge n-1$, note that the cycle $(1,2, \cdots, n)$ cannot be decomposed into a product of fewer than $n-1$ transpositions. 

These above considerations also give a conceptually simple\footnote{If one is given an arbitrary permutation, computing the number of disjoint cycles is hard, requiring $O(n) \sim O(2^K)$ computations.} formula for the complexity. Let $c(g)$ be the number of disjoint cycles in the decomposition of $g$. Then the complexity of $g$ is
\eqn{C(g) = n-c(g).}

To further probe the geometry, consider a random walk on the space. This is similar to probing a curved manifold by studying diffusion on the manifold or by putting a scalar field on it. There is also a complexity interpretation of the random walk that makes this problem interesting in its own right; namely, consider a random circuit model, where at each time step we choose a simple gate $s\in S$ at random and append it to the end of the existing circuit $g \to s g$. This is equivalent to a random walk on the permutation group where all steps of unit distance have equal probability! Random walks on finite groups have been intensely studied in the math literature, see, e.g., \cite{diaconis02, saloff} for a review.

\begin{figure}[H]
	\centering
	\includegraphics[scale=0.75, trim = 0cm 0cm 0cm 0cm, clip=true]{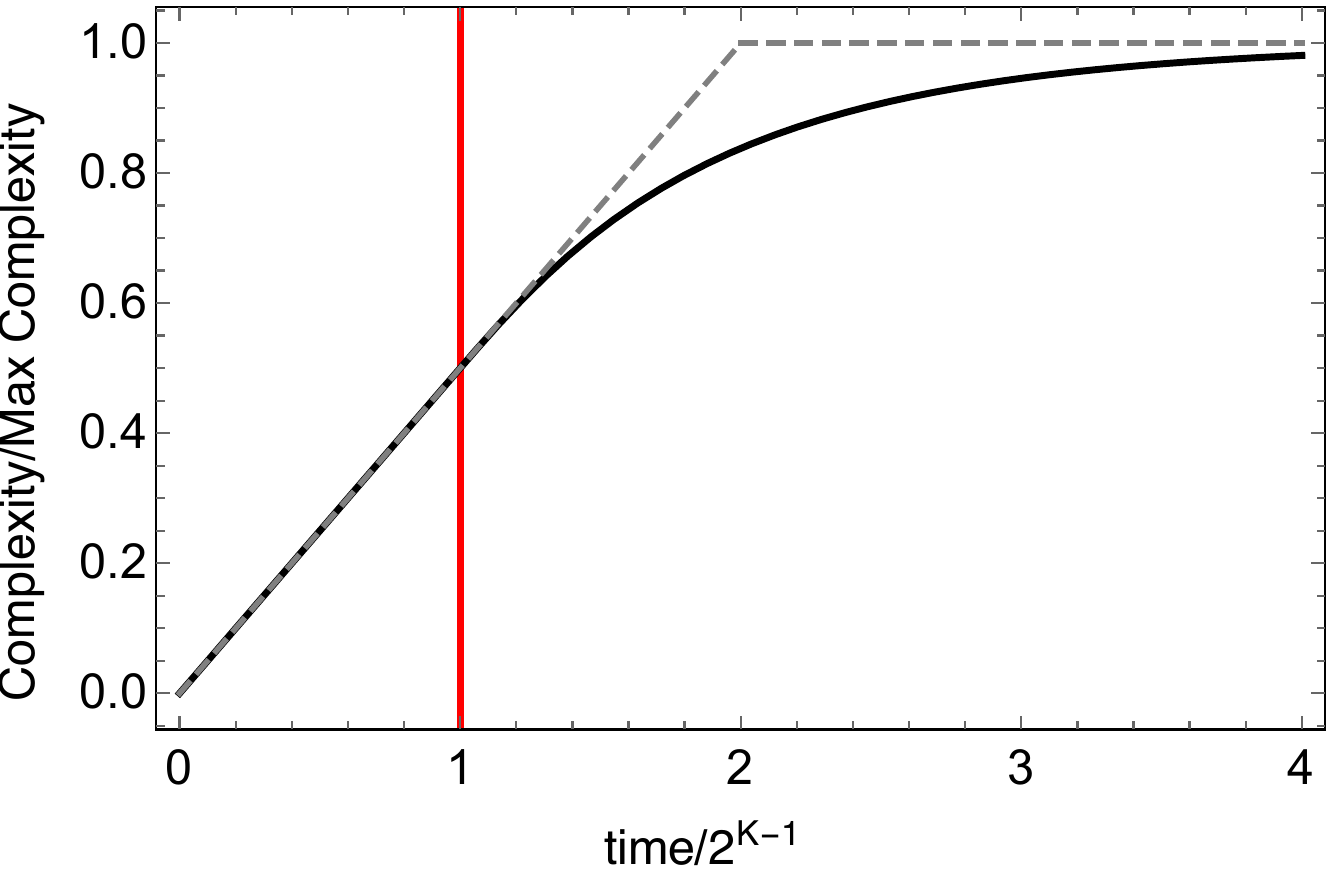}
	\caption{\label{favg} The average complexity as a function of time in the random circuit model (solid black) defined in the text. In the large $K$ limit, there is a phase transition at $t=2^{K-1}$ (red), where the second derivative is discontinuous. The dashed curve is what was conjectured in, e.g., figure 1 of \cite{brown17}. }
\end{figure}

In the limit where $n = 2^K \to \infty$, the results of \cite{berdur06} give a formula for the average complexity $C$ as a function of time, with the initial condition that the walk starts at the origin:
\eqn{C/n = 1 - \sum_{k=1} {1 \over \tau} {k^{k-2} \over k!} (\tau e^{-\tau})^k,}
where $\tau = 2t/n$. (In fact, \cite{berdur06} characterizes the fluctuations of this curve). For $t< n/2$, it can be shown that the complexity grows at exactly unit speed $C = t$. For $t >n/2$, it can be shown that $C < t$. In particular, the second derivative of $C$ is discontinuous at $t =n/2$. We plot this function in figure \ref{favg}.

This result can be understood intuitively as follows \cite{berdur06}. Given any group element $g$, consider a decomposition of into random transpositions. Associate to a $g$ a graph (certainly {\it not} to be confused with the Cayley graph), where the vertices are elements of the integers $\{1, \cdots,n\}$. An edge between $a_1$ and $a_k$ exists if there is a transposition $(a_1, a_k)$.\footnote{We allow multiple edges between vertices to take care of the unlikely scenario where a transposition occurs more than once.} As $g$ executes a random walk starting from the origin, this graph will become less and less sparse. Bigger and bigger connected components will form; finally at $t = n/2$, there is an Erdos-Renyi phase transition (in the large $n$ limit) where the largest connected component becomes macroscopic (e.g., spans a finite fraction of the vertices). After this time, there will be a finite probability for a transposition to land in this connected component and break up the connected component. Since the complexity is $n-c$, where $c$ is roughly the number of components, breaking up a component decreases the complexity.

The behavior of this random walk has geometric implications. A particularly interesting one is due to a theorem of \cite{ber06} about the thickness of triangles in the Cayley graph geometry as $n \to \infty$. (We refer the reader to the original paper for a more precise statement). 

{\it Theorem}: Let $T$ be a triangle formed by the origin and two points sampled independently from the hitting distribution on the sphere of radius $a 2^K$ for $0<a<1$. If $a<1/4$ and $K \to \infty$, then with probability 1, $T$ is $\delta$-thin with $\delta \sim O(1)$, whereas $\delta \sim O(2^K)$ for $a > 1/4$. 

Roughly speaking, this theorem says that on small scales, triangles are $\delta$-thin, whereas on scales comparable to the diameter of the group $D= 2^K$, there is a breakdown of $\delta$-hyperbolicity. The random walk grows linearly with distance when the geometry is hyperbolic; at late times it ``feels'' the compactness of the space and slows down in its growth.

We have emphasized the similarities between our toy model and the complexity geometry of the unitary group. However, this model also has features which we do not expect of a complexity geometry defined by generic $k$-local set of unitaries.

The most important difference is the fast scrambling time. As discussed in \cite{brown16}, to get a scrambling time of order $K \log K$, we need a curvature length scale $\delta \sim K$. Our above results show that $\delta \sim 1$. Note however, that if we rescale the distances to account for the fact that each simple gate has $k$-local complexity of order $K$, then we also find a curvature scale $\delta \sim K$. Despite this agreement in curvature scale, our model will not exhibit the switchback effect. To understand why, recall that the switchback effect involves studying the complexity of the operator $\tilde{h} = g h g\inv$, where $h$ is a simple operator, and $h$ is obtained from a random circuit of length $t$. The failure of $h$ to commute with gates in $g$ leads to an epidemic, such that the complexity of $\tilde{h}$ grows exponentially with $t$ until a time defined as the scrambling time. However, one can in fact show that in our toy model, $C(\tilde{h}) = C(h)$ for any possible $g$ since complexity is only a function of the conjugacy class of the element. 

More geometrically, if we choose $D$ directions at random on our Cayley graph, where $D \ll 2^{K}$, we will see with high probability that the subgraph generated by these directions will look like a hypercube of dimension $D$ in these directions. The large number of loops\footnote{I thank Lenny Susskind for discussions related to this point.} prevent any switchback effect.

It is worth noting that these differences between the complexity geometry are not artifacts of the finite group we have chosen, but the generating set. If we had chosen $S$ to be the set of all $k$-local reversible classical gates, then we would have found a curvature scale $\sim K$, as well as a switch back effect. One argument for the curvature scale is the following. A cruder way\footnote{Using the commutator to measure curvature is like using squares to probe the geometry instead of triangles.} to measure curvature is to use the commutator $[g,h] = g h g\inv h\inv$. If the commutator vanishes, then locally the graph looks like a square lattice, which we consider to be flat. If we draw $g, h$ from a sphere of radius of $R$ surrounding the identity, the length scale $R$ on which the commutator becomes non-trivial is a measure of the curvature scale. If our gates are $k$-local, then $g$ and $h$ will commute with a probability $\mathcal{O}(1/K)$. Hence $\delta \sim K$.  

Before proceeding, let us comment on the complexity of a discrete analog of time-independent Hamiltonian evolution (as opposed to a random circuit). In a finite group, the order $T$ of any element, defined to be the smallest integer such that $g^T = 1$ is finite. $T$ is the discrete analog of the recurrence time in quantum mechanics. It is the time it takes for the complexity to fluctuate back to 0. It can be shown that if $g$ is uniformly chosen from $S_n$ then the average recurrence time as $n\to\infty$ is \cite{goh}: 
\eqn{T \sim \exp \lp c \lp {n\over \log n }\rp^2 \rp,}
where $c^2 = 8 \int_0^\infty \log \log (e/(1-e^{-t}) dt   \approx 8.94$. This timescale is doubly exponentially large in $K$, in agreement with \cite{brown17}.

\subsection{More general Cayley geometry}
So far we have explored the complexity geometry of a particular Cayley graph. In this example, it was possible to explicitly compute the average complexity as a function of time. For a more generic Cayley graph, such a computation seems daunting. However, some properties of the graph can lead to interesting constraints on complexity. In this section, we will briefly discuss two such properties.

The first property is related to the statement that complexity geometry is negatively curved. Besides $\delta$-hyperbolicity, another property of hyperbolic geometry is that the area of a sphere is proportional to its volume. An expander graph shares this property; one definition of an $c$-expander graph $G$ is that for any subset of vertices $v$, the number of neighbors $|N(v)|$ satisfies
\eqn{|N(v)| >  c|v|f,}
where $f = 1-|v|/|G|$ is just a finite volume correction factor. There are many explicit examples of Cayley graphs that are expanders. In fact, a random Cayley graph will do the trick! More precisely, the Alon-Roichman theorem \cite{alon1994random} states that there exists some function $s(c)$ such that for any group of order $n$ and for a random generating set $S$ of size $s(c) \log n$, the corresponding Cayley graph is a $c$-expander with a probability that tends to 1 as $n \to \infty$. 
A famous and explicit family of Cayley graphs associated to the groups $G_p = PSL(2,\mathbb{Z}_p)$ for prime numbers $p$ have been constructed \cite{lubotzky1988ramanujan} that have asymptotically optimal expansion properties. These groups have faithful unitary representations of dimension $p$, so we can think of them as acting on quantum system of Hilbert space dimension $\ge p$.


The second property is the existence of a dynamical phase transition known as the {\it cutoff}. Let $p(t)$ be the probability distribution $p(t)$ that characterizes the system, and let $\mu$ be the equilibrium distribution. For a Cayley graph, we take $p(t)$ to be a probability distribution over the vertices of the graph generated by a random walk starting at $1$. The random walk exhibits a cutoff if the $L_1$ distance between the probability distributions
\eqn{ D_N(t) ={1\over 2} \sum_i |p_i(t)-\mu_i| }
jumps from its maximal value of $1$ to zero at the cutoff time $t_c$. Said more carefully, 
\eqn{\lim_{N \to \infty} D_N(t_c(1+\epsilon)) = \Theta(\epsilon).}
The canonical example of a cutoff is card shuffling \cite{aldous1986shuffling, saloff2004random}, see \cite{diaconis02} for a fun exposition. With $n$ cards, a card configuration can be identified with an element of the permutation group of order $n!$. Card shuffling can then be modeled as a random walk on the Cayley graph with a particular choice of $S$. 
As $n \to \infty$ cards, there is a sharp transition between the deck being not shuffled and shuffled. If the permutation group is generated by transpositions as in our example, the cutoff time is ${1\over2} n \log n$. Note that this is a different timescale than the phase transition that separates linear and non-linear growth of the complexity. The cutoff occurs at a time when the difference between the complexity of the random walk and the equilibrium complexity is within the fluctuations of the complexity at equilibrium, which is later due to the late time sub-linear growth. In general, the cutoff will give us an estimate for the time it takes for the complexity of a random circuit to reach its maximum. 

There are a large number of Cayley graphs which are known or conjectured to exhibit cutoff (see \cite{saloff2004random} for a review). For example, many different generating sets of the permutation group or the alternating group exhibit cutoff. An interesting future direction would be to prove or disprove the existence of a cutoff on the alternating group $A(2^{K})$ generated by $k$-local classical reversible gates. If one could show that the cutoff time is as small as possible (e.g., of order the diameter of the graph), this would prove that the complexity grows linearly and then has a sharp turnover.

\section{Discussion}

In this section, we discuss how the Cayley graph geometry is similar and different to the continuum case. 

An interesting point about the Cayley graph geometry is that we did not have to choose penalty factors for complex directions; the distance to a highly complicated operator is automatically determined by just a choice of simple gates $S$. In the case of a Lie group, the analog of the condition that $S$ be a generating set is that a choice of simple Hamiltonians (elements of the Lie algebra) generate the full Lie algebra of the Lie group under commutators. This is in contrast to the approach of \cite{nielsen2006quantum, brown17}, where additional penalty factors are chosen.

Let us explore the analog of this in the finite group setting by considering a generalization of the Cayley graph. We can take all elements in $G$ to be the generating set, but associate to each pair $(g,g\inv)$ some penalty factor $I_g$. Then the distance $d$ along some path $\mathcal{P}$ would be 
\eqn{d(\mathcal{P}) = \sum_{g\in \mathcal{P}} I_g.}
We define the distance between two points as the minimum distance over all paths. First note that if $I_g = 1$ for all elements, we get the discrete analog of the Fubini-Study or bi-invariant metric, where all elements are a distance of $O(1)$ from all others.

Now imagine that we take $I_S = 1$ on some set $S$ and let $I_g \to \infty$ for $g \notin S$. Naively, we might worry that the distances between points are diverging. However, if $S$ is a generating set, we get nothing but the geometry of the Cayley graph. In fact, we do not need to take $I_{S'} \to \infty$ to get the Cayley graph geometry. Once $I_{S'}$ is larger than the diameter of the Cayley graph, the distance between any two points on the graph is independent of $I_{S'}$, since a route through the Cayley graph will always be preferred. 
Specializing to the permutation group generated by transpositions, we only need to require that $I_g > n-c(g)$ by equation (2.1). For the reversible computations generated by $k$-local gates, we can define the size of an operator \cite{roberts2018operator} $s(g)$ to be the number of qubits on which $g$ acts nontrivially. Then $I(g)$ just needs to be larger than the maximum complexity of any operator of size $s(g)$, which is exponential in $s$. All these examples show that the Cayley graph geometry is in a sense universal: as long as the penalty factors on elements not in $S$ are large, we recover the Cayley graph geometry independent of detailed choices we make for the penalty factors.

An interesting question is whether this universality has an analog in the continuum version of the complexity geometry. This issue will be discussed thoroughly in an upcoming publication.

As a final speculation, we would like to discuss the formation of a firewall in holography. The failure of the complexity to grow linearly with time has been advocated as a signature of a firewall \cite{susskind-firewall}. In the toy model of complexity depicted in figure \ref{favg}, there is a genuine Erdos-Renyi phase transition separating linear and sub-linear growth. More generally, we anticipate that the complexity geometry should display cutoff phenomena, which may or may not coincide with the phase transition in the growth of complexity. 

It is interesting that these phase transitions, if they persist in more realistic models, have a candidate holographic interpretation as the formation of a firewall \cite{susskind-firewall}. Even some of the details are qualitatively consistent with this picture; for example, before the transition, the complexity curve of a single instance of the random walk will be linear (reflecting a smooth dual geometry) with exponentially suppressed fluctuations, whereas after the transition, the complexity curve will become jagged for individual instances. Understanding to what extent phase transitions in the complexity growth is universal could be a fruitful future direction.

\section*{Acknowledgements}
I thank Adam Brown, Juan Maldacena, Dan Roberts, Douglas Stanford, Lenny Susskind, Victor Wang, and Ying Zhao for helpful discussions and encouragement. I am supported in part by an NDSEG fellowship.
\begin{appendix}

\section{Appendix: Complexity of random permutations}
First we prove a classical theorem due to Toffoli that says that it is only possible to create even permutations starting from local gates:\\
{\bf Theorem} \cite{toffoli1980reversible}: Any circuit acting on $K$ bits, consisting of $k$-local gates, where $k < K$ computes an even permutation. Hence we only generate the group $A_{2^K} \subset S_{2^K}$.

To prove this theorem, it suffices to show this is true for a single $k$-local gate, since a product of even permutations is even. Since $k<K$, there must be at least one bit which does not act on. Without loss of generality, assume it is the last bit. Now decompose the permutation into cycles. Since the last bit is fixed, each cycle comes in pairs of the form $(a_1 1, a_21, a_3 1, \cdots, a_n 1) (a_1 0, a_2 0, \cdots,a_n 0 )$. Therefore the permutation is even.

It can be shown that any even permutaiton can be generated by $3$-local classical gates. In fact, we only need the gates CNOT, NOT, TOFFOLI. We can convert any odd permutation into an even one by adding one bit.

{\bf Theorem}: The complexity of a pair of transpositions $(a,b)(c,d)$ is linear in $K$. Since a generic transposition acts on all bits, the complexity must be at least linear in $K$. To show an upper bound, we must construct a circuit. This is done explicitly in \cite{shende}; they give an upper bound on the complexity of $16(K-5) + (5K-2)= 21K-82$ using the above universal gate set.

\section{Subtleties about the group}
In our definition of universal classical reversible gates, we only require that our gates generate the alternating group, the set of all even elements of the permutation group. An alternative definition of the alternating group is the set of permutations with determinant $+1$. 

This is related to a subtlety in the definition of quantum complexity. Sometimes it is only required that a set of universal gates generate $SU(2^K)$, instead of $U(2^K)$. The motivation in the quantum case is that the overall phase of the unitary does not matter. The permutation group $S_{2^K}$ is a subgroup of $U(2^K)$ but not $SU(2^K)$. 

We would also like to point out that if the overall phase is really treated as unphysical, we should think of the complexity geometry not being defined on $SU(2^K)$ but on the projective group $PU(D,\C) = U(D,\C)/U(1)$. Using this definition, the distance between $U$ and $\omega U$, where $\omega^{2^K}= 1$ is always zero, whereas it could be non-zero if the complexity geometry were defined on $SU(2^K)$. 

If we consider the group $PU(D)$ instead of $SU(D)$, the correct bi-invariant metric is the Fubini-Study metric:
\eqn{ds^2 = { \Tr dU^\dagger dU \over  \Tr U^\dagger U} - { \Tr dU^\dagger U  \Tr U^\dagger dU \over \lp \Tr U^\dagger U\rp^2}.\label{fs}}
Of course, for unitary elements the denominators are trivial, but if we wished analytically continue to Euclidean time, $U$ would be an element of $PGL(D,\C) = GL(D,\C)/\C^*$ and the denominators would be non-trivial.

\section{Classical complexity geometry}
Since the classical permutation group is discrete, one might think that a smooth complexity geometry is a quantum feature. However, if we allow for probabilistic operations on classical bit strings, there is a classical analog of complexity geometry. Such probabilistic operations naturally occur if the logic gates in a classical circuit occasionally produce an error. (For example, an AND gate could function as an OR gate with probability $\epsilon$). Such faulty circuit elements are represented by Markov matrices. These are matrices whose entries are conditional probabilities. They act not on a wavefunctions but on a probabilities. The analog of the bi-invariant metric on the space of unitaries is the Fisher metric on the Markov matrices.

The set of matrices that are both Markov and unitary are the permutation matrices. The discrete model is thus the limit of two inequivalent physical systems: one is classical but stochastic, the other is quantum. One could wonder whether classical (but probabilistic) complexity plays any role in Euclidean AdS/CFT.

In quantum mechanics, we can think of a unitary matrix as defining some operation we could act on a state with. However, we could also think of it as defining some maximally entangled state. There is a close analog in classical probability. We can think of a permutation matrix as defining a joint probability distribution on two copies of our bit strings. The marginal distributions for each copy is completely random, but if we know the conditional entropy between the two sides vanishes (they are perfectly correlated).

\end{appendix}

\mciteSetMidEndSepPunct{}{\ifmciteBstWouldAddEndPunct.\else\fi}{\relax}
\bibliographystyle{utphys}
\bibliography{rw}

\end{document}